\documentclass[aps,prl,twocolumn,amsmath,amssymb,nofootinbib,superscriptaddress,floatfix]{revtex4}

\usepackage{amsmath} 
\usepackage{amssymb} 
\usepackage{amsthm}
\usepackage[pdftex]{color} 
\usepackage{graphicx}
\usepackage{dcolumn} 
\usepackage{bm} 
\usepackage{longtable}
\usepackage{ulem} 
\newcommand{\bs}[1]{{\boldsymbol{#1}}}

\begin{document}

\title{How to Measure the Quantum Geometry of Bloch Bands}

\author{Titus Neupert} \affiliation{ Condensed Matter Theory Group,
  Paul Scherrer Institute, CH-5232 Villigen PSI, Switzerland }

\author{Claudio Chamon} \affiliation{ Physics Department, Boston
  University, Boston, Massachusetts 02215, USA }
            
\author{Christopher Mudry} \affiliation{ Condensed Matter Theory
  Group, Paul Scherrer Institute, CH-5232 Villigen PSI, Switzerland }

\date{\today}

\begin{abstract}
Single-particle states in electronic Bloch bands form a Riemannian
manifold whose geometric properties are described by two gauge
invariant tensors, one being symmetric the other being
antisymmetric, that can be combined into the so-called Fubini-Study
metric tensor of the projective Hilbert space.  The latter directly
controls the Hall conductivity.  Here we show that the symmetric
part of the Fubini-Study metric tensor also has measurable
consequences by demonstrating that it enters the current noise
spectrum. In particular, we show that 
a non-vanishing equilibrium current noise spectrum
at zero temperature is unavoidable whenever Wannier states
have non-zero minimum spread, 
the latter being quantifiable by
the symmetric part of the Fubini-Study metric tensor. We
    illustrate our results by
three examples: (1) atomic layers of
  hexagonal boron nitride, (2) graphene, and (3) the surface
  states of three-dimensional topological insulators when gaped by
  magnetic dopants.
\end{abstract}

\maketitle 

The connection between geometry and quantum mechanics was
explored systematically during the 80's when it was realized that the
projective space of normalized quantum states can be equipped with
a distance, making it a Riemannian manifold, and a symplectic form,
making it a Kaehlerian manifold%
~\cite{Provost80,Thouless82,Berry84,Heslot85,Anadan90,Pati91,probecriticality}.
Berry famously showed that a quantum state acquires a measurable
phase factor of purely geometric origin during a cyclic
adiabatic evolution~\cite{Berry84}, i.e.,
he showed that the
symplectic form (the Berry curvature)
on the projective space of normalized quantum states
is proportional to the phase acquired by a state under 
an infinitesimal adiabatic cycle. 
As this description applies to any subspace of the
projective Hilbert space that smoothly depends on a set 
of external parameters, it is also of relevance to non-interacting Bloch
bands~\cite{Thouless82,Marzari97,Souza00,Ma10,Matsuura10,Roy12}, 
where the crystal momentum
parametrizes the manifold of quantum states.

Most known measurable consequences of the quantum geometry of band
insulators are limited to the Berry curvature. For example,
the integral over the Brillouin zone (BZ) of the Berry curvature
is quantized and proportional to the Hall conductivity of a band
insulator~\cite{Thouless82}.  
It also enters the semiclassical equations of motion of electronic
wavepackets~\cite{Shindou05}.
Here, we show that the quantum geometric tensor, also known as 
the Fubini-Study metric tensor of complex projective spaces
in the mathematical literature~\cite{Fubini03},
is an observable that can be measured via the current noise spectrum 
of a band insulator.

We consider the family of single-particle Bloch Hamiltonians
\begin{equation}
\mathcal{H}(\bm{k}):=
\sum_{a=1}^{N}
\varepsilon^{\,}_{a}(\bm{k})\,
|u^{\,}_{a}(\bm{k})\rangle
\langle u^{\,}_{a}(\bm{k})|
\label{eq: def mathcal H(k)}
\end{equation}
labeled by the momentum $\bs{k}$ from the $d$-dimensional BZ of volume 
$\Omega^{\,}_{\mathrm{BZ}}$ acting on the Hilbert
space $\mathbb{C}^{N}$.
For any momentum $\bs{k}\in\mathrm{BZ}$,
the single-particle Bloch eigenstates
$|u^{\,}_{a}(\bm{k})\rangle$ labeled by the band index $a=1,\cdots,N$
are orthonormal $N$-dimensional complex-valued vectors that span
the Hilbert space $\mathbb{C}^{N}$. The projective Hilbert space
$\mathbb{CP}^{N-1}$ is obtained from $\mathbb{C}^{N}$ by identifying
any two vectors $\bm{v}$ and $\bm{w}$ from $\mathbb{C}^{N}$ related
to each other by the multiplication of a non-vanishing complex number.

We first review how the Fubini-Study metric tensor 
on the projective Hilbert space $\mathbb{CP}^{N-1}$
arises. To this end, we define the normalized single-particle state
\begin{equation}
|\Psi(\bm{k})\rangle:=
\sum_{\tilde{a}=1}^{\widetilde{N}}
c^{\,}_{\tilde{a}}(\bm{k})\,
|u^{\,}_{\tilde{a}}(\bs{k})\rangle,
\qquad
\sum_{\tilde{a}=1}^{\widetilde{N}}
|c^{\,}_{\tilde{a}}(\bm{k})|^{2}=1,
\end{equation}
whereby we assume that the first $\widetilde{N}$ bands
are separated from the remaining $N-\widetilde{N}$ bands
by a spectral gap. 
We want to compute the infinitesimal increment
\begin{equation}
(\mathrm{d}s)^{2}:=
\sum_{\mu,\nu=1}^{d}
\langle\partial^{\,}_{\mu}\Psi(\bm{k})|
\partial^{\,}_{\nu}\Psi(\bm{k})\rangle\,
\mathrm{d}k^{\mu}\,
\mathrm{d}k^{\nu}
\end{equation}
under the (adiabatic) assumption that the state
$|\Psi(\bm{k}+\mathrm{d}\bm{k})\rangle$ has no overlap
with any of the bands above the gap
[geometrically, we parallel transport
the state $|\Psi(\bm{k})\rangle$ to the state
$|\Psi(\bm{k}+\mathrm{d}\bm{k})\rangle$]. 
One finds that~\cite{Ma10}
\begin{subequations}
\begin{equation}
(\mathrm{d}s)^{2}=
\sum_{\mu,\nu=1}^{d}
\left(
\sum_{\tilde{a},\tilde{b}=1}^{\widetilde{N}}
c^{*}_{\tilde{a}}(\bm{k})\,
Q^{\tilde{a}\tilde{b}}_{\mu\nu}(\bm{k})\,
c^{\,}_{\tilde{b}}(\bm{k})
\right)
\mathrm{d}k^{\mu}\,
\mathrm{d}k^{\nu}.
\end{equation}
For any pair $\mu,\nu=1,\cdots,\widetilde{N}$,
the non-Abelian Fubini-Study metric tensor 
$Q^{\,}_{\mu\nu}(\bm{k})$ 
on the complex projective space $\mathbb{CP}^{N-1}$ 
is here the $\widetilde{N}\times\widetilde{N}$
Hermitean matrix
\begin{equation}
Q^{\,}_{\mu\nu}(\bm{k}):=
g^{\,}_{\mu\nu}(\bm{k})
+
\mathrm{i}
\omega^{\,}_{\mu\nu}(\bm{k}).
\end{equation}
It can be decomposed additively in a unique way into
the Hermitean $\widetilde{N}\times\widetilde{N}$ matrix
$g^{\,}_{\mu\nu}(\bm{k})$
with the components
\begin{equation}
\begin{split}
g^{\tilde{a}\tilde{b}}_{\mu\nu}(\bs{k}):=&\,
\frac{1}{2}
\Big[ 
\langle
\partial^{\,}_{\mu}u^{\,}_{\tilde{a}}(\bs{k})|
\partial^{\,}_{\nu} u^{\,}_{\tilde{b}}(\bs{k})\rangle
-
\sum_{\tilde{c}=1}^{\widetilde{N}}
A^{\tilde{a}\tilde{c}}_{\mu}(\bs{k})\,
A^{\tilde{c}\tilde{b}}_{\nu}(\bs{k}) 
\\
&\,
+
(\mu\leftrightarrow\nu) 
\Big]
\end{split}
\end{equation}
and the Hermitean $\widetilde{N}\times\widetilde{N}$ matrix
$\omega^{\,}_{\mu\nu}(\bm{k})$ with the components
\begin{equation}
\begin{split}
\omega^{\tilde{a}\tilde{b}}_{\mu\nu}(\bs{k}):=
\frac{1}{2}
F^{\tilde{a}\tilde{b}}_{\mu\nu}.
\end{split}
\end{equation}
We have made use of the non-Abelian Berry connection 
\begin{equation}
A^{ab}_{\mu}(\bs{k}):=
-
\mathrm{i}
\langle
u^{\,}_a(\bs{k})|\partial^{\,}_{\mu} u^{\,}_b(\bs{k})
\rangle
\end{equation}
together with its non-Abelian Berry field strength
\begin{equation}
F^{ab}_{\mu\nu}:=
\partial^{\,}_{\mu}
A^{ab}_{\nu}(\bs{k})
-
\partial^{\,}_{\nu}
A^{ab}_{\mu}(\bs{k})
+
\mathrm{i}
\left[
A^{\,}_{\mu}(\bs{k}),
A^{\,}_{\nu}(\bs{k})
\right]^{ab},
\end{equation}
\end{subequations}
for any $a,b=1,\cdots,N$ that we have projected
onto the $\widetilde{N}$ lower bands by restricting the
band labels to $\tilde{a},\tilde{b}=1,\cdots,\widetilde{N}$.

In the following, we shall consider the case of a band insulator
with $\widetilde{N}=1$, i.e., with a single band $a=1$ filled and all
other bands $a=2,\cdots, N$ empty and separated by an energy
gap from the lowest band~\cite{footnote:manybands}.

The current noise spectrum is the Fourier transform of the
current-current correlation
function~\cite{Lesovik97,Blanter00,Gavish03,footnote}
\begin{subequations}
\begin{equation}
S^{\,}_{\mu\nu}(\omega):=
\int\limits\mathrm{d}t\,e^{-\mathrm{i}\,\omega\,t} 
\langle
0|J^{\,}_{\mu}(0)J^{\,}_{\nu}(t)|0
\rangle
\label{eq: def S(omega)}
\end{equation}
for any pair $\mu,\nu=1,\cdots,d$.
The insulating non-interacting many-body ground state 
is here denoted $|0\rangle$. It has
the lowest band $a=1$ filled and all other bands empty.
The time-dependence of the current operator is
\begin{equation}
\bs{J}(t):=e^{\mathrm{i}\,Ht}\bs{J}e^{-\mathrm{i}\,Ht}.
\label{eq: definition current as fct t}
\end{equation}
The initial value of the current operator
\begin{equation}
\bs{J}\equiv\bm{J}(0):=\mathrm{i}[H,\bs{X}]
\label{eq: definition initial current}
\end{equation}
is proportional to the commutator between the 
non-interacting Hamiltonian $H$ 
with the single-particle representation%
~(\ref{eq: def mathcal H(k)})
and the position operator $\bm{X}$ 
with the single-particle representation
\begin{equation}
\bm{X}=
\int\limits_{\mathrm{BZ}}\frac{\mathrm{d}^{d}\bs{k}}{\Omega^{\,}_{\mathrm{BZ}}}\,
|u^{\,}_{a}(\bs{k})\rangle 
\left[
-\mathrm{i}\delta^{ab}\bm{\partial}
+
\bm{A}^{ab}(\bs{k})
\right] 
\langle u^{\,}_{b}(\bs{k})|
\label{eq: def position operator}
\end{equation} 
\end{subequations}
(the sum over the repeated band labels $a,b=1,\cdots, N$ is implicit
and $\Omega^{\,}_{\mathrm{BZ}}$ denotes the volume of the BZ).

To proceed with the derivation of our main result,
we assume that the current in the ground state vanishes,
\begin{equation}
\langle 0|\bm{J}(t)|0\rangle =0.
\label{eq: 0 J 0 =0}
\end{equation}
With the help of the resolution of the identity
\begin{equation}
\openone=
\sum_{n=0}^{\infty}|n\rangle\langle n|=
|0\rangle\langle0|
+
\sum_{m=1}^{\infty}|m\rangle\langle m|,
\end{equation}
where $|m\rangle$ denotes any one of the many-body eigenstates 
except for the ground state with the many-body eigenenergy
$E^{\,}_{m}$ measured relative to the ground state eigenenergy, 
we can rewrite Eq.~\eqref{eq: def S(omega)} using 
Eqs.~(\ref{eq: definition current as fct t}),
\eqref{eq: definition initial current},
and~(\ref{eq: 0 J 0 =0}) as
\begin{equation}
S^{\,}_{\mu\nu}(\omega)=\sum_m \int\limits\mathrm{d}t\,
e^{-\mathrm{i}\,(\omega-E^{\,}_m) t} \langle 0|J^{\,}_{\mu}
|m\rangle\langle m|J^{\,}_{\nu}|0\rangle.
\end{equation}
As $\bs{J}$ is a single-particle operator, 
it can only create particle-hole excitations above the
ground state with energy
$E^{\,}_{m}=
\varepsilon^{\,}_{a}(\bs{k})-\varepsilon^{\,}_{1}(\bs{k}')$,
where $m=(a,\bs{k},\bs{k}')$, $a>1$,
$\bs{k},\bs{k}'\in\mathrm{BZ}$. 
Thus,
\begin{equation}
\begin{split}
S^{\,}_{\mu\nu}(\omega)=&\,
\int\limits_{\mathrm{BZ}}\frac{\mathrm{d}^{d}\bs{k}}{\Omega^{\,}_{\mathrm{BZ}}}
\int\limits_{\mathrm{BZ}}\frac{\mathrm{d}^{d}\bs{k}'}{\Omega^{\,}_{\mathrm{BZ}}} 
\sum_{a>1}
\int\limits\mathrm{d}t\,
e^{-\mathrm{i}\,[\omega-\varepsilon^{\,}_{a}(\bs{k})+\varepsilon^{\,}_{1}(\bs{k}')]
  t} 
\\ 
&
\times
\left[\varepsilon^{\,}_{a}(\bs{k})-\varepsilon^{\,}_{1}(\bs{k}')\right]^{2}
\langle 0|X^{\,}_{\mu} |m\rangle\langle m|X^{\,}_{\nu}|0\rangle
\\
=&\,
2\pi\omega^{2}
\int\limits_{\mathrm{BZ}}\frac{\mathrm{d}^{d}\bs{k}}{\Omega^{\,}_{\mathrm{BZ}}}
\int\limits_{\mathrm{BZ}}\frac{\mathrm{d}^{d}\bs{k}'}{\Omega^{\,}_{\mathrm{BZ}}} 
\sum_{a>1}
\\ 
&
\times
\delta\Big(\omega-\varepsilon^{\,}_{a}(\bs{k})+\varepsilon^{\,}_{1}(\bs{k}')\Big)
\langle 0|X^{\,}_{\mu} |m\rangle\langle m|X^{\,}_{\nu}|0\rangle.
\end{split}
\end{equation}
By inspection of Eq.~\eqref{eq: def position operator}, we observe that
the position operator decomposes additively into a band-diagonal 
but momentum-off-diagonal part (the derivative in momentum space) 
and a band-non-diagonal but momentum-diagonal part 
(the non-Abelian Berry connection).  Only the latter contributes to the
matrix elements $\langle 0|X^{\,}_{\mu} |m\rangle$, since the electron has
to be excited to an upper band $a>1$. Hence,
\begin{equation}
\begin{split}
S^{\,}_{\mu\nu}(\omega)=&\,
2\pi\omega^{2}
\int\limits_{\mathrm{BZ}}\frac{\mathrm{d}^{d}\bs{k}}{\Omega^{\,}_{\mathrm{BZ}}} 
\sum_{a>1}
\delta
\left[
\omega-\varepsilon^{\,}_{a}(\bs{k})+\varepsilon^{\,}_{1}(\bs{k})
\right]
\\ 
&\times 
A^{1a}_{\mu}(\bs{k})A^{a1}_{\nu}(\bs{k}).
\end{split}
\label{eq: intermediate result Smunu} 
\end{equation}

To relate Eq.~\eqref{eq: intermediate result Smunu} to the quantum
geometric tensor $Q^{\,}_{\mu\nu}$, we would like to resort to the
following manipulation (we need the single-particle
resolution of the identity to establish the first equality)
\begin{equation}
\begin{split}
\sum_{a>1} A^{1a}_{\mu}(\bs{k})A^{a1}_{\nu}(\bs{k})=&\,
A^{11}_{\mu}(\bs{k})A^{11}_{\nu}(\bs{k}) 
-
\langle
\partial^{\,}_{\mu}u^{\,}_{1}(\bs{k})|
\partial^{\,}_{\nu}u^{\,}_{1}(\bs{k})
\rangle
\\ 
=&\,
-
Q^{11}_{\mu\nu}(\bs{k}).
\end{split}
\end{equation} 
However, in general we cannot perform the summation over $a>1$ in
Eq.~\eqref{eq: intermediate result Smunu}, for the energies
$\varepsilon^{\,}_{a}(\bs{k})$ also depend on $a=1,\cdots,N$, 
so that energetics and quantum geometry combine in $S^{\,}_{\mu\nu}(\omega)$. 
We will now discuss two ways to distill the contribution from the quantum
geometry.

On the one hand, we have the sum rule
\begin{equation}
\mathcal{S}^{\,}_{\mu\nu}:=
\int\limits
\frac{\mathrm{d}\omega}{2\pi}
\frac{S^{\,}_{\mu\nu}(\omega)}{\omega^{2}}=
-
\int\limits_{\mathrm{BZ}}\frac{\mathrm{d}^{d}\bs{k}}{\Omega^{\,}_{\mathrm{BZ}}}\,
Q^{11}_{\mu\nu}(\bs{k})
\label{eq: sum rule main result}
\end{equation}
that relates the frequency integral of the current noise spectrum
divided by $\omega^{2}$ 
to the integral of the quantum geometric tensor over the BZ.
On the other hand, when $N=2$, i.e., for exactly two bands,
\begin{equation}
\begin{split}
S^{\,}_{\mu\nu}(\omega)=&\,
-2\pi\omega^{2}
\int\limits_{\mathrm{BZ}}\frac{\mathrm{d}^{d}\bs{k}}{\Omega^{\,}_{\mathrm{BZ}}}
\delta
\left[\omega-\varepsilon^{\,}_{2}(\bs{k})+\varepsilon^{\,}_{1}(\bs{k})
\right]
\,
Q^{11}_{\mu\nu}(\bs{k}) ,
\end{split}
\label{eq: Two bands Smunu main result} 
\end{equation}
so that $S^{\,}_{\mu\nu}(\omega)/\omega^{2}$ equals the integral of the
quantum geometric tensor over the region in momentum space where the
\textit{direct} band gap equals $\omega$.  The reduction to a two-band
model with $a=1,2$ is justified when the orbital character of the
bands $a\geq3$ is sufficiently different from the band $a=1$, such
that
\begin{equation}
[P^{\,}_{1},P^{\,}_{a}]\approx0,\qquad a\geq3,
\end{equation}
holds, where 
$P^{\,}_{a}:=
\int\limits\mathrm{d}^{d}\bs{k}\,\Omega^{-1}_{\mathrm{BZ}}\,
|u^{\,}_{a}(\bs{k})\rangle\langle u^{\,}_{a}(\bs{k})|$ is the projector on
the single-particle states of the band $a=1,2,\cdots$. In this case
$A^{1a}_{\mu}(\bs{k})$ is negligible for $a\geq 3$ and so do its
contributions to Eq.~\eqref{eq: intermediate result Smunu}.

Equations~(\ref{eq: sum rule main result})
and~(\ref{eq: Two bands Smunu main result})
establish a connection between the quantum geometry of the Bloch
states and the physically measurable current noise spectrum. On the
one hand, the frequency dependence of the noise can reveal information
on the Fubini-Study metric tensor. 
On the other hand, in multi-orbital systems or
materials with spin-orbit interactions (in which the quantum metric tensor
is generically non-trivial), there are interesting structures in the
noise spectra even at equilibrium. To illustrate the latter case, 
we consider three examples that can be realized experimentally.

\begin{figure}[t]
\begin{center}
\includegraphics[width=84mm]{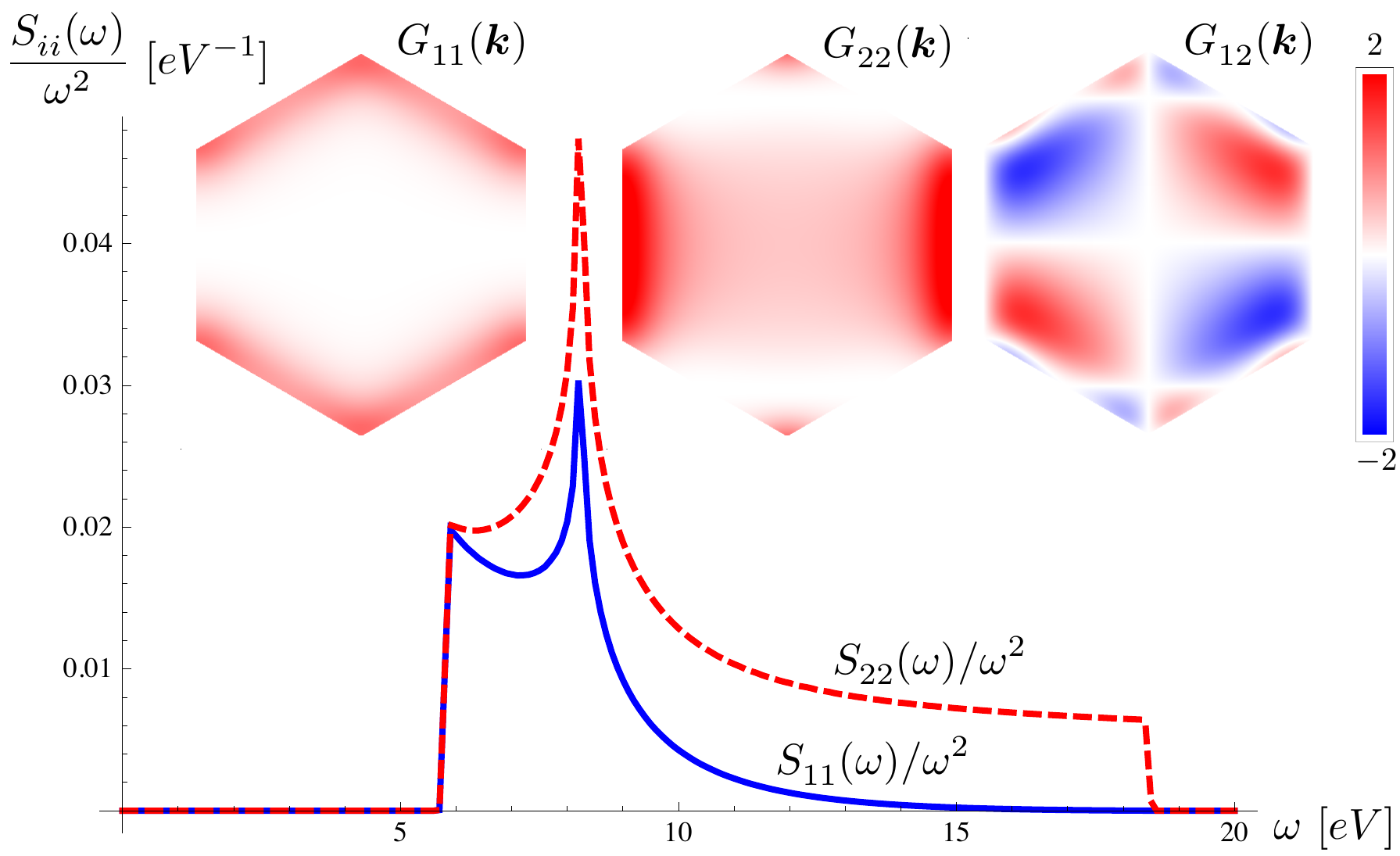}
\caption{(Color online) Current noise spectrum computed using
Eq.~\eqref{eq: Two bands Smunu main result}  
and Hamiltonian~\eqref{eq: Hamiltonian BN} 
for longitudinal currents in boron nitride.  It is a direct
measure for the Fubini-Study metric tensor times the density of states
averaged over equal-energy contours in the BZ. The largest
contributions stem from the massive Dirac cones directly above the
band gap of $\omega=2\mu_{\rm s}=5.8\,\mathrm{eV}$ and from the van Hove
singularities at $\omega=2\sqrt{t^{2}+\mu_{\rm s}^{2}}=8.23\,\mathrm{eV}$. The
anisotropy is attributed to the fact that the honeycomb lattice
lacks a four-fold rotational symmetry.  Insets show the distribution
of the components of Fubini-Study metric tensor over the BZ.  
        }
\label{fig: noise spectrum for BN}
\end{center}
\end{figure}

\textit{Example 1} --- We consider atomic layers of hexagonal boron
nitride. In the tight-binding approximation, the electronic structure
is described by the gaped Hamiltonian
\begin{equation}
\mathcal{H}^{\mathrm{BN}}(\bs{k}) :=
\begin{pmatrix}
\mu_{\rm s}& -t\, \gamma(\bs{k}) \\ -t\, \gamma^*(\bs{k})&-\mu_{\rm s}
\end{pmatrix},
\label{eq: Hamiltonian BN}
\end{equation}
where
$\gamma(\bs{k})=1+e^{-\mathrm{i}\bs{a}^{\,}_1\cdot\bs{k}}+e^{-\mathrm{i}\bs{a}^{\,}_2\cdot\bs{k}}$,
$t=2.92\,\mathrm{eV}$ is the nearest neighbor hopping,
$\mu_{\rm s}=2.90\,\mathrm{eV}$ is the difference in chemical potential
between boron and nitrogen sites, and
$\bs{a}^{\,}_1=(\sqrt{3},3)^{\mathsf{T}}/2$,
$\bs{a}^{\,}_2=(-\sqrt{3},3)^{\mathsf{T}}/2$ are the primitive lattice
vectors (in units of the atomic spacing).
Hamiltonian~\eqref{eq: Hamiltonian BN} has two bands
separated by the band gap $2\mu_{\rm s}$. While neither of these bands has a
nontrivial topological attribute, they still represent a nontrivial
quantum geometry. The off-diagonal components of the quantum geometric
tensor are nonzero, but average to zero along equal energy contours in
momentum space, so that $S^{\,}_{12}(\omega)=0$ according to
Eq.~\eqref{eq: Two bands Smunu main result}.  
As far as the Berry curvature is concerned, 
this averaging is a consequence of time-reversal symmetry.
On the other hand, $S^{\,}_{\mu\mu}(\omega)$, $\mu=1,2$, is nonzero and shown
in Fig.~\ref{fig: noise spectrum for BN}.  Finally,
$\mathcal{S}^{\,}_{\mu\mu}$, $\mu=1,2$, defined in Eq.%
~\eqref{eq: sum rule main result}, 
are given by 
$\mathcal{S}^{\,}_{11}=-1.54\,\mathfrak{a}^{2}$ 
and
$\mathcal{S}^{\,}_{22}=-3.56\,\mathfrak{a}^{2}$,
where the lattice spacing $\mathfrak{a}$ has been reinstated.

\textit{Example 2} --- 
The limit $\mu^{\,}_{\mathrm{s}}\to0$
in Eq.~(\ref{eq: Hamiltonian BN})
delivers a tight-binding two-band approximation to the 
bands of graphene. When the chemical potential is tuned to
the charge-neutral point, graphene realizes a quantum critical point
characterized by a density of states that scales linearly
with the deviation in energy away from the charge-neutral point.
Correspondingly, the diagonal entries 
$S^{\,}_{\mu\mu}(\omega)$ 
with $\mu=1,2$
scale linearly with $\omega$ as $\omega\to0$. 
It follows that 
$\mathcal{S}^{\,}_{\mu\mu}$ with $\mu=1,2$
are logarithmically divergent due to the
critical nature of the Bloch states at the
charge-neutral point.

\textit{Example 3} --- We consider a single species of massive Dirac
elections in $d=2$ dimensions, as a model for the surface states of
the three-dimensional topological insulator Bi$_2$Se$_3$~\cite{Xia09},
when doped with ferromagnetically ordered ions. The Hamiltonian is
given by
\begin{equation}
\mathcal{H}^{\mathrm{TI}}(\bs{k}) :=
\begin{pmatrix}
m& v(k^{\,}_2+\mathrm{i}k^{\,}_1) \\ v(k^{\,}_2-\mathrm{i}k^{\,}_1)&-m
\end{pmatrix},
\label{eq: Hamiltonian TI}
\end{equation}
where $v$ is the Fermi velocity and $m$ is the magnetization out of
the plane of the surface.  Hamiltonian~\eqref{eq: Hamiltonian TI} has
two bands with energies $\pm\varepsilon(\bs{k})$,
$\varepsilon(\bs{k}):=\sqrt{v^{2}\bs{k}^{2}+m^{2}}$, 
separated by the band gap $2m$.  
We regulate the theory with a high-energy cut-off 
$\Lambda\gg m$ such that $\Omega^{\,}_{\mathrm{BZ}}=\pi(\Lambda/v)^{2}$. 
The quantum geometric tensor reads
\begin{equation}
\begin{small}
Q^{11}=\frac{v^{2}}{4\varepsilon(\bs{k})^4}
\begin{pmatrix}
2k^{2}\sin^{2} \varphi+2m^{2}&
\mathrm{i}m\,\varepsilon(\bs{k})-k^{2}\sin2\varphi
\\ -\mathrm{i}m\,\varepsilon(\bs{k})-k^{2}\sin2\varphi
&2k^{2}\cos^{2}\varphi+2m^{2}
\end{pmatrix}.
\end{small}
\end{equation}
Here, we used the parametrization
$v\bs{k}=k(\cos\varphi,\sin\varphi)^{\mathsf{T}}$.
For the current noise spectrum, we obtain ($\mu=1,2$)
\begin{subequations}
\begin{eqnarray}
S^{\,}_{\mu\mu}(\omega) &=& 
-\frac{\pi
v^{2}}{\Lambda^{2}}
\left(\omega+\frac{4m^{2}}{\omega}\right)
\Theta\left(\frac{\omega}{2m}-1\right),
\\ 
S^{\,}_{12}(\omega) &=& 
-\mathrm{i}
\frac{2\pi v^{2}m}{\Lambda^{2}}
\Theta\left(\frac{\omega}{2m}-1\right),
\end{eqnarray}
\label{eq: result Dirac S}
\end{subequations}
while
\begin{equation}
\mathcal{S}^{\,}_{12}= 
-\mathrm{i}\frac{2\pi}{\Omega^{\,}_{\mathrm{BZ}}}
\left[\frac{1}{2}
+
\mathcal{O}\left(\frac{m}{\Lambda}\right)\right],
\label{eq: Dirac integrated noise}
\end{equation}
reveals that the Chern number of a single species of Dirac fermions is
$1/2$.  In contrast, $\Omega^{\,}_{\mathrm{BZ}}\times\mathcal{S}^{\,}_{\mu\mu}$, 
$\mu=1,2$, is
logarithmically divergent for $\Lambda\to\infty$.  One might wonder
whether bulk states, that have not been considered here, will spoil
these results. If fact, the results are valid as long as $\omega$ in
Eq.~\eqref{eq: result Dirac S} and $\Lambda$ in
Eq.~\eqref{eq: Dirac integrated noise} are much
smaller than the bulk energy gap.

The results for $S^{\,}_{\mu\nu}(\omega)$ and
$\mathcal{S}^{\,}_{\mu\nu}$ obtained in these three examples
illustrate how the quantum geometry is manifest in the noise. 
Notice that in example 2 there is no current noise at equilibrium conditions, 
while in examples 1 and 3 there is necessarily noise even at equilibrium.


We have shown that the tensor
$\mathcal{S}$ defined by the first equality
of Eq.~\eqref{eq: sum rule main result}
is connected to the current noise spectrum
by the second equality
of Eq.~\eqref{eq: sum rule main result}.
We are going to provide yet two complementary interpretations 
for this tensor.

Two physical quantities that are revealed in 
$\mathcal{S}$ are the minimum spread of
Wannier states and the Hall conductivity of a Bloch band. 
As shown by Marzari and Vanderbilt~\cite{Marzari97}, the spread of the
Wannier states can be broken into two positive definite contributions
$\Omega^{\,}_{I}+\tilde\Omega$, one of which ($\Omega^{\,}_{I}$) 
is gauge invariant and is tied to the trace of the 
quantum geometric tensor. It turns out that
\begin{equation}
\Omega^{\,}_{I}=
\int\limits_{\mathrm{BZ}}\frac{\mathrm{d}^{2}\bs{k}}{\Omega^{\,}_{\mathrm{BZ}}}\,
{\rm tr}\,g^{11}(\bs{k})=
-\mathrm{tr}\,\mathcal{S}.
\label{eq:Mazari_Vanderbilt_spread_noise}
\end{equation}
Remarkably, current noise is present even at equilibrium
for any band insulator in which either
multi-orbital or spin-orbit coupling causes the Wannier states to
spread.
Furthermore, the imaginary part of $\mathcal{S}$ is proportional to the
Hall conductivity $\sigma^{\mathrm{H}}_{\mu\nu}$ with $\mu\neq\nu=1,2$
of the lower band $a=1$ 
\begin{equation}
\sigma^{\mathrm{H}}_{\mu\nu}=
2\pi\,\frac{e^2}{h}
\int\limits_{\mathrm{BZ}}\frac{\mathrm{d}^{2}\bs{k}}{\Omega^{\,}_{\mathrm{BZ}}}\,
F^{11}_{\mu\nu}(\bs{k})=
-2\pi\, \frac{e^2}{h}\mathrm{Im}\,\mathcal{S}_{\mu\nu}.
\end{equation}
The fluctuation-dissipation theorem relates $S^{\,}_{\mu\nu}(\omega)$
to the frequency-resolved inter-band Kubo conductivity 
$\sigma^{\,}_{\mu\nu}(\omega)$
per unit volume $V$. At zero temperature,
\begin{equation}
\sigma^{\,}_{\mu\nu}(\omega)=
\frac{\mathrm{i}}{2\pi\,V}\,
\int
\frac{
\mathrm{d}\omega^{\prime}
     }
     {
\omega^{\prime}
     }
\frac{
S^{\,}_{\mu\nu}(+\omega^{\prime})
-
S^{\,}_{\nu\mu}(-\omega^{\prime})
     }
     {
\omega
-
\omega^{\prime}
+
\mathrm{i}0^{+}
       }.
\end{equation} 
This implies sum rules
relating the Fubini-Study metric $Q^{\,}_{\mu\nu}(\bm{k})$
and $\sigma^{\,}_{\mu\nu}(\omega)$%
~\cite{footnote:Souza00}.

The integrated noise spectrum $\mathcal{S}$ can also be
interpreted as the action of the $\mathbb{CP}^{N-1}$
nonlinear sigma model (NL$\sigma$M)~\cite{cpnNLSM}.
To this end, we note that the orthonormal eigenstates of the
$N\times N$ Bloch Hamiltonian~(\ref{eq: def mathcal H(k)})
can be represented as points $\bm{z}(\bm{k})$
on the surface of the unit sphere
$S^{2N-1}$.
Any two pair of points
$\bm{z}(\bm{k})$ and $\bm{z}(\bm{k}')$
from $S^{2N-1}$
differing by a phase are not distinct, i.e.,
it is the projective space
$\mathbb{CP}^{N-1}$
that realizes physical states.
We can interpret $\mathbb{CP}^{N-1}$
as a $(2N-2)$-dimensional real Riemannian manifold,
with the ``angular'' coordinates 
$\phi^{\,}_{\mathsf{a}}(\bs{k})$, 
$\mathsf{a}=1,\cdots,2N-2$.
In this parametrization, the Fubini-Study metric tensor
decomposes into the symmetric
\begin{equation}
g^{11}_{\mu\nu}=
\partial^{\,}_{\mu}\phi^{\,}_{\mathsf{a}}\,
\mathcal{G}^{\mathsf{a}\mathsf{b}}(\bs{\phi})\, 
\partial^{\,}_{\nu}\phi^{\,}_{\mathsf{b}}=
+g^{11}_{\nu\mu}
\end{equation}
and antisymmetric 
\begin{equation}
F^{11}_{\mu\nu}=
\partial^{\,}_{\mu}\phi^{\,}_{\mathsf{a}}\,
\mathcal{F}^{\mathsf{a}\mathsf{b}}(\bs{\phi})\, 
\partial^{\,}_{\nu}\phi^{\,}_{\mathsf{b}}=
-F^{11}_{\nu\mu}
\label{eq: metric and curvature in terms of phi} 
\end{equation} 
tensors, respectively (summation over repeated 
$\mathsf{a},\mathsf{b}=1,\cdots,2N-2$ is implied).
In $d=2$ dimensions, given the flat 
Euclidean metric tensor
$\delta^{\mu\nu}=+\delta^{\nu\mu}$
and the
Levi-Civita antisymmetric tensor
$\epsilon^{\mu\nu}=-\epsilon^{\nu\mu}$,
we can write (summation over repeated indices implied)
\begin{subequations}
\begin{equation}
\delta^{\mu\nu}\mathcal{S}^{\,}_{\mu\nu}[\bs{\phi}]=
-
\int\limits_{\mathrm{BZ}}\,
\frac{\mathrm{d}^{2}\bs{k}}{\Omega^{\,}_{\mathrm{BZ}}}
\partial^{\,}_{\mu}\phi^{\,}_{\mathsf{a}}\,
\mathcal{G}^{\mathsf{a}\mathsf{b}}(\bs{\phi})\,
\partial^{\,}_{\mu}\phi^{\,}_{\mathsf{b}},
\label{eq: NLSM 1}
\end{equation}
and
\begin{equation}
\epsilon^{\mu\nu}\mathcal{S}^{\,}_{\mu\nu}[\bs{\phi}]=
-
\frac{\mathrm{i}}{2}
\int\limits_{\mathrm{BZ}}\,
\frac{\mathrm{d}^{2}\bs{k}}{\Omega^{\,}_{\mathrm{BZ}}}
\epsilon^{\mu\nu}\,
\partial^{\,}_{\mu}\phi^{\,}_{\mathsf{a}}\,
\mathcal{F}^{\mathsf{a}\mathsf{b}}(\bs{\phi})\,
\partial^{\,}_{\nu}\phi^{\,}_{\mathsf{b}}.
\label{eq: NLSM 2}
\end{equation}
\end{subequations}
Equation~\eqref{eq: NLSM 1} 
is the kinetic term in the action of the
$\mathbb{CP}^{N-1}$ NL$\sigma$M
in two-dimensional Euclidean space.
Equation~\eqref{eq: NLSM 2} 
is the Wess-Zumino term of the $\mathbb{CP}^{N-1}$ NL$\sigma$M%
~\cite{Witten83}. 
The quantization and with it the topological character of the Wess-Zumino
term is guaranteed by the quantization of the first Chern number.
Measuring all the components of the
tensor $\mathcal{S}^{\,}_{\mu\nu}$ can thus be viewed as
measuring the action of the $\mathbb{CP}^{N-1}$ NL$\sigma$M
augmented by a topological term
with the field configuration $\bs{\phi}(\bs{k})$ 
that is dictated by the Bloch Hamiltonian.

Finally, we point out that the Fubini-Study metric tensor
enters the algebra obeyed by the
single-particle position operator%
~(\ref{eq: def position operator}),
which we denote by $\widetilde{\bm{X}}$
after projection onto the $\widetilde{N}$ lower bands,
according to
\begin{equation}
\left\langle
u^{\,}_{\tilde{a}}(\bm{k})
\left|
\widetilde{X}^{\,}_{\mu}\,
\widetilde{X}^{\,}_{\nu}\,
\right|
u^{\,}_{\tilde{b}}(\bm{k})
\right\rangle=
Q^{\tilde{a}\tilde{b}}_{\mu\nu}(\bm{k})
\end{equation}
for any pair $\mu,\nu=1,\cdots,d$ and for any pair
$\tilde{a},\tilde{b}=1,\cdots,\widetilde{N}$
from the lower bands.
Furthermore, the Fubini-Study metric tensor
determines the algebra obeyed by the Fourier components
of projected density operators
\begin{equation}
\widetilde{\rho}(\bs{q}):=
\int\limits_{\mathrm{BZ}}
\frac{\mathrm{d}^{d}\bs{k}}{\Omega^{\,}_{\mathrm{BZ}}}\,
|u^{\,}_{\tilde{a}}(\bs{k})\rangle
\langle u^{\,}_{\tilde{a}}(\bs{k})|u^{\,}_{\tilde{b}}(\bs{k}+\bs{q})\rangle
\langle u^{\,}_{\tilde{b}}(\bs{k}+\bs{q})|,
\label{eq: density operator}
\end{equation}
that reads in the limit of long wavelength, 
i.e., to second order in the momenta 
$\bs{q},\bs{q}'\in\mathrm{BZ}$,
\begin{equation}
\begin{small}
\widetilde{\rho}(\bs{q})\widetilde{\rho}(\bs{q}')
-
\widetilde{\rho}(\bs{q}+\bs{q}')=
q^{\,}_{\mu} q^{\prime}_{\nu}
\int\limits_{\mathrm{BZ}}\frac{\mathrm{d}^d\bs{k}}{\Omega^{\,}_{\mathrm{BZ}}}
|u^{\,}_{\tilde{a}}(\bs{k})\rangle
Q^{\tilde{a}\tilde{b}}_{\mu\nu}(\bs{k})
\langle u^{\,}_{\tilde{b}}(\bs{k})|.
\end{small}
\end{equation}

In conclusion, we showed that the quantum geometric tensor of band insulators
is related to a measurable quantity, the current noise spectrum. 
We also introduced a frequency-weighted integral of the noise spectrum
that can be physically interpreted as the minimal spread of Wannier orbitals
and takes the form of the action of the $\mathbb{CP}^{N-1}$ NL$\sigma$M 
augmented by a topological term.

This work was supported in part by DOE Grant DEFG02-06ER46316 and by
the Swiss National Science Foundation.


\begin{thebibliography}{99}
\bibitem{footnote:manybands} 
We choose a single occupied band out of
simplicity.  The generalization of our results to more than one
occupied band is straight-forward.
\bibitem{footnote} 
The noise spectrum can be defined using either the
  anti-symmetrized (see Ref.~\cite{Blanter00}) or the
  non-antisymmetrized current-current correlator (see
  Refs.~\cite{Lesovik97,Gavish03}), depending on the measurement
  scheme. Here we concentrate on the latter. We stress, however, that
  one can still recover the quantum metric tensor from measurements of
  the former.
\bibitem{footnote:Souza00}
A sum rule relating $g^{\,}_{\mu\mu}(\bm{k})$ with $\mu=1,2$
to $\mathrm{Re}\,\sigma^{\,}_{\mu\mu}(\omega)$
was found in Ref.~\cite{Souza00}.
\end{thebibliography}
\end{document}